\documentclass[epj,final]{svjour}

\usepackage{latexsym}
\usepackage{url}
\usepackage{amsfonts}

\RequirePackage{graphicx}

\begin{document}
\title{On CCC-predicted concentric low-variance circles in the CMB sky}
\author{V. G. Gurzadyan\inst{1} \and R. Penrose\inst{2}
}                     
%
%
\institute{Alikhanian National Laboratory and Yerevan State University, Yerevan, Armenia \and 
Mathematical Institute, 24-29 St Giles', Oxford OX1 3LB, U.K}
\date{Received: date / Revised version: date}
%

\abstract{
A new analysis of the CMB, using WMAP data, supports earlier indications 
of non-Gaussian features of concentric circles of low temperature variance. Conformal cyclic cosmology (CCC) 
predicts such features from supermassive black-hole encounters in an \textit{aeon} preceding our Big Bang. The significance of \textit{individual} low-variance circles in the true data has been disputed; yet a recent independent analysis  has confirmed CCC's expectation that CMB circles have a non-Gaussian temperature distribution. Here we examine \textit{concentric sets} of low-variance circular rings in the WMAP data, finding a highly \textit{non-isotropic} distribution. A new ``sky-twist" procedure, directly analysing WMAP data, without appeal to simulations, shows that the prevalence of these concentric sets depends on the rings being \textit{circular}, rather than even slightly elliptical, numbers dropping off dramatically with increasing ellipticity. This is consistent with CCC's expectations; so also is the crucial fact that whereas some of the rings' radii are found to reach around $15^\circ$, 
none exceed $20^\circ$. The non-isotropic distribution of the concentric sets may be linked to previously known anomalous and non-Gaussian CMB features. 
} 

\maketitle
\section*{Introduction}

A key prediction of conformal cyclic cosmology (CCC; \cite{Penrose2010}, subsection 3.6, \cite{Gurzadyan2010a}) is the presence of families of concentric low-variance circular rings in the CMB (cosmic microwave background). We present a new analysis, indicating 
the presence of such families in the actual CMB sky. Using a simple but novel``sky-twist" procedure, we argue that these are not statistical artifacts, by comparing the numbers of such families of circular rings with corresponding numbers of families of {\it elliptically distorted} rings in the CMB, according to WMAP's (Wilkinson Microwave Anisotropy Probe) 7-year data. We find a dramatic fall-off with increasing ellipticity, for concentric sets of at least 3 low-variance rings, where for rings that are clearly visibly elliptical, their 
numbers drop by a factor of 10 or more, to numbers comparable with those previously found \cite{Gurzadyan2011}, for concentric {\it circular} rings in conventional simulated skies, with CMB power spectrum incorporated.  Thus we point out the particular role of the circularity of these features in the CMB, as brought out by the geometrically revealing sky-twist test. We regard this present account as providing a preliminary indication of these CCC-predicted signals being present in the CMB, though we do not rule out more conventional explanations of these features. When Planck satellite's data is publicly released, this may provide more definitive and revealing information.

We examine 10885 points in the CMB sky, as potential centres, which are defined specifically as centres of 2D-balls of equal numbers of pixels, starting from the equator and shifted by $1.5^\circ$ in latitude $b$, including the longitude $l=0$ at each latitude row (a sample being depicted in Fig. 2 of \cite{Gurzadyan2010a}, excluding the region $|b|<20^\circ$ containing the galactic disc. Our use of the term ``low-variance circle" (or ``low-variance ring") in the CMB, is as follows. Having ascertained that the dips in temperature variance of rings in the WMAP data drop away sharply when the angular radius increases beyond $16^\circ$, we calculate the temperature variance over circular rings (of annular width $0.5^\circ$ and of angular radius taken in steps of $0.5^\circ$ in the range $2.5^\circ-16^\circ$). For each of these centres in turn, the temperature variance - a standard deviation $\sigma$, in units $\mu$K - is calculated for each of the rings with this centre, of radius within $2.5^\circ-16^\circ$. A ring is considered to be of ``low variance" if, among all those with 
this same centre, its temperature variance is found to dip by at least $15\mu$K below the average variance for that particular centre. If successive rings satisfy this criterion, they are considered as a single (slightly wider) ring, rings being considered separate if the variance rises above threshold, for some ring between them. Whether or not parts of rings within $|b|<20^\circ$ are counted, or if the KQ85 mask \cite{Gold} is used instead, makes no significant difference to the main conclusions. For definiteness we have not counted ring portions (necessarily less than semi-circular) within $|b|<20^\circ$, though we include all centres with $|b|\geq20^\circ$ in our explicit counts, as discussed below. 

In earlier notes \cite{Gurzadyan2010a,Gurzadyan2010b,Gurzadyan2011} we argued that the WMAP data indicates some evidence of concentric sets of low-variance circular rings in the CMB, this being in accordance with a prediction of \textit{conformal cyclic cosmology} (abbreviated CCC; see \cite{Penrose2006,Penrose2010}). This prediction of CCC was described, in essence, in \cite{Penrose2010}, but the criterion of finding circles of distinctively \textit{low variance}, was not suggested (nor was the expectation of their appearance in concentric sets explicitly mentioned), but it was pointed out that CCC should predict some circles with a distinctively higher temperature and some lower. We appear to find evidence of this also, but more importantly a recent independent statistical analysis by \cite{Meissner} has established the non-Gaussian presence of such circles with 99.7\% confidence. We consider that the criterion of low variance is a clearer diagnostic of CCC, as it is hard to see how such an effect, when significant, could arise from some other kind of disturbance. In Appendix B, we demonstrate that the CCC-predicted circles should indeed be distinguished by low variance, but we do not attempt to estimate CCC's specific prediction of how low, as this depends upon various presently unknown factors. However, it is certainly an expectation of CCC that a major factor in the CMB temperature variations at the smallest scale would be low-variance circles that have arisen from supermassive black-hole encounters in the previous aeon. Accordingly, these contributions ought to be found just about at the level of the temperature variations actually seen. Although the expectation that the circles ought frequently to be in concentric sets was not explicitly pointed out in \cite{Penrose2010}, it is a clear implication of the discussion put forward there and was made explicit in \cite{Gurzadyan2010a}.

In fact \textit{individual} low-variance circles do also appear - in similar numbers to those found in the true WMAP data - in simulated data incorporating WMAP's CMB power spectrum plus overlapped WMAP's noise; see \cite{Moss,Wehus,Hajian}. We subsequently pointed out \cite{Gurzadyan2010b,Gurzadyan2011} that the sought-for features of temperature variance, if present in the actual CMB data, necessarily influence the power spectrum itself, arguing that the appearance of individual low-variance circles in simulations is enhanced by WMAP data's power spectrum. The finding by \cite{Meissner} of high significance non-Gaussian circular rings of larger or smaller average temperature in the CMB, is complementary to those of this paper, as we are concerned here with families of \textit{multiple} concentric rings of \textit{low variance}. The specific non-isotropic distribution of such rings that we find can be linked to other signatures of anomalous or non-Gaussian features revealed in the CMB (e.g. \cite{Starkman}).

In Section 2, we demonstrate that for centres of at least 3, or of at least 4, concentric rings in the actual WMAP data, the large numbers that we find (352 for $\geq$ 3 rings, 56 for $\geq$ 4 rings) depend \textit{crucially} on the individual rings being \textit{circular}; for when we apply the same search for concentric rings which depart from exact circularity, distorted to various degrees of ellipticity, we find a dramatic fall-off in numbers, as the elliptic eccentricity increases. For the greatest degree of ellipticity that we examine $(S=\pm 80)$, for which the major/minor axis ratio is around 2, the fall-off reaches the same kind of low value that we had previously found for circular rings in simulated skies (incorporating the empirical power spectrum, beaming, synfast, R=0.02, and WMAP's noise). Moreover, we find that the \textit{distribution} of the centres of these concentric sets of circles in the WMAP data is very far from random. This is not what is expected for a Gaussian sky - such as would be anticipated on the conventional inflationary picture of CMB temperature variations resulting from random quantum fluctuations of an inflaton field. The inhomogeneity of this distribution is interesting in its own right, well worthy of further detailed study. From the point of view of CCC, this inhomogeneity, though not actually anticipated by the theory, can be explained by a gross non-uniformity in the distribution and sizes of galactic clusters in the aeon prior to ours.

In Section 1, we outline the essentials of CCC and the reasons for expecting \textit{low-variance} circles (see Appendix B) with tendency to appear in concentric sets. We also indicate some features of the average temperatures over such circles, anticipated by CCC. Technical details of the mathematical structure of CCC are supplied in Appendix A. Section 2 exhibits our findings on centres of 3 or more low variance circles in the CMB sky: in a region covering the whole sky excluding the galactic strip $|b|<20^\circ$ there are 352 centres of at least three low-variance circles in the WMAP data, and 56 centres of at least four such circles. We show, by means of a simple ``sky-twist" (similar to that of \cite{Penrose2010}, p.218, in a slightly different context) applied to this multiple-circle search, that these numbers drop dramatically when applied to slightly elliptically deformed rings, with fewer centres found the larger the deviation from exact circularity - a clear indication that the features we find in the true CMB are indeed basically concentric circles, as anticipated be CCC, rather than distorted circular shapes. Another noteworthy point is that the true data reveal circles of up to $15^\circ$ radius but we find none exceeding $20^\circ$, as anticipated by CCC \cite{Tod2011,Nelson}.

\section{Outline of CCC and its relevant implications}

As described in \cite{Penrose2010}, CCC proposes that our present picture of an indefinitely expanding universe, starting with the Big Bang and ending with an exponential expansion, in accordance with the presence of a positive cosmological constant $\Lambda$ (but where no early inflationary period is taken to have occurred), is but one \textit{aeon} of an indefinitely continuing succession of qualitatively similar such aeons, the conformal infinity of each joining smoothly to the conformally stretched big bang\footnote{The capitalized ``Big Bang" refers to our aeon's initial state; ``big bang", to that of an aeon generally.} of the next. Thus, it is proposed that there was an aeon prior to ours whose exponentially expanding remote future joins to our Big Bang across a 3-manifold ${\cal X}$ (the crossover 3-surface), the resulting 4-manifold being conformally smooth across ${\cal X}$. Whereas CCC does not incorporate inflation as such, the ultimate exponential expansion of the previous aeon would play a role in several respects similar to that of inflation, but this exponential expansion occurs \textit{prior} to the Big Bang, rather than immediately following it (compare \cite{Gasperini}). The main observational distinction between CCC and conventional inflation is that whereas in the latter the initial seeds of inhomogeneity are taken to be randomly occurring quantum fluctuations, in CCC inhomogeneities result from various causes, in the previous aeon, all subject to the exponential (largely self-similar) ultimate expansion of that aeon governed, for the most part, by \textit{classical} partial differential equations.

A key motivation for CCC comes from a need to incorporate the second law of thermodynamics (abbreviated $2^{nd}$ Law), in the form that we find it, with the gravitational degrees of freedom enormously suppressed in the initial state. This can be expressed in the ``Weyl curvature hypothesis" (see \cite{Penrose2004}, subsection 28.8), a particularly elegant form of which is that proposed by \cite{Tod2003}, which demands that the Big Bang can be conformally stretched to become a smooth initial conformal boundary to our current aeon. CCC adopts Tod's proposal, but goes further by requiring our aeon to be the conformal continuation of a preceding aeon that resembles our own in broad terms. By far the major increase in entropy throughout the evolution of each aeon occurs through the formation of huge black holes in galactic centres, these ultimately all disappearing via Hawking evaporation after enormous time periods of up to $\sim 10^{100}$ years. This is associated with a huge loss of phase-space volume\footnote{A key element of CCC's consistency with the $2^{nd}$ Law is \cite{Hawking1976} information loss via final black-hole evaporation. More recently \cite{Hawking2005} reversed his position, but CCC takes his original argument as more cogent. Information loss implies not a violation of the $2^{nd}$ Law but that the entropy definition undergoes a shift when applied to the entire universe, its zero being re-set when a black hole disappears (\cite{Penrose2010}, subsection 3.4).}, causing the entropy definition to be ``renormalized", so that it ultimately becomes re-set at a (relatively) extremely low value somewhat before the start of the next aeon.

According to CCC, the dynamics of black holes in the previous aeon has strong observational implications for our own aeon, where we suppose that this previous aeon was indeed generally similar to ours. It would therefore have contained galaxies with huge central black holes, in clusters that remain bound for the most part, despite the cosmological exponential expansion. We must expect occasional encounters between such supermassive black holes when galaxies within a single cluster encounter one another, often leading to mergers of the holes. This would have resulted in enormous bursts of energy in the form of gravitational radiation, each such burst carrying off a sizable proportion (perhaps a few per cent) of the total mass-energy of the merging pair of holes. On the very broad scale that we are considering, such bursts would have been effectively impulsive and normally largely spherically symmetric in overall intensity\footnote{In exceptional situations a black-hole collision can result in a single hole with a large proper motion \cite{Redmount,Merritt}. In such particular circumstances, the gravitational wave impulse would carry away appreciable momentum, and have significantly non-isotropic overall intensity.
}, where the energy would, in effect, be carried outwards from the source along a very narrow space-time region following the future light cone of the source point. In the conformal picture of Fig.1, we see that this cone expands until reaching the 3-surface ${\cal X}$ representing, according to CCC, both the future conformal infinity of the previous aeon and the conformally stretched Big Bang of our own. (This diagram is drawn, according to standard conventions of conformal diagrams, with light rays tilted at $45^\circ$ to the vertical, where we take the horizontal dimension to represent Euclidean 3-geometry of constant cosmic time.) The cone continues into our aeon, as drawn in Fig.1, until reaching our last scattering 3-surface ${\cal L}$. This energy impulse would emerge from ${\cal X}$ not as gravitational radiation, but as an impulse imparted to the initial form of the dark matter that would, according to CCC's equations, be necessarily created at ${\cal X}$ (Appendix A).

In CCC, the essential processes carrying information across ${\cal X}$ come from entirely deterministic classical partial differential equations, with \textit{no quantum input} of the kind currently suggested for producing temperature variations in the CMB, such as quantum fluctuations in an inflaton field. According to CCC, these temperature fluctuations are all initiated by astrophysical processes in the previous aeon. Thus, information is able to propagate from aeon to aeon, in CCC, despite the enormous temperatures encountered in the early stages of each big bang. The basic relevant mathematical equations for this are provided in Appendix A (for more details see \cite{Penrose2010}, Appendix B). These equations tell us, for example, that electromagnetic radiation (and magnetic fields) can propagate directly through ${\cal X}$, from aeon to aeon.

The reason that gravitational-wave impulses from previous-aeon black-hole encounters transform to impulsive disturbances in the initial form of dark matter in our present aeon comes from the behaviour of the conformal factors describing the relation between our aeon and the previous one. Einstein's physical metric tensor $\check {\mathit{g}}$ of our aeon is related to the corresponding physical metric tensor $\hat {\mathit{g}}$ for the previous aeon via an auxiliary third metric tensor $g$, smooth across ${\cal X}$, covering the join between the two. Conformal factors $\Omega$ and $\omega$ define the relations between these metrics according to $\hat{g}=\Omega^2g$ and $\check{g}=\omega^2g$. We have $\Omega\rightarrow\infty$ and $\omega\rightarrow 0$ at ${\cal X}$, and CCC also requires the ``reciprocal hypothesis" $\omega=-\Omega^{-1}$. There are further conditions (Appendix A) ensuring essential uniqueness of $g$ and continuation across ${\cal X}$. Whereas the conformal factor $\Omega$, appropriate to the earlier aeon may be viewed, there, as a ``phantom field", having no direct physical content, $\Omega$ acquires a \textit{physical} role in the subsequent aeon, naturally interpreted as the initial form of newly created dark matter, taking up the asymptotic degrees of freedom of the gravitational field of the earlier aeon, whereas gravitational degrees of freedom, \textit{as such}, become hugely suppressed in this subsequent aeon.

Equations governing this transformation are described in Appendix A, but we provide a brief outline here. The final exponential expansion of the earlier aeon leads to its Weyl conformal curvature tensor \textbf{\textit{C}} vanishing at ${\cal X}$ \cite{Penrose1984,Friedrich}. The assumed conformal smoothness across ${\cal X}$ ensures \textbf{\textit{C}}$=0$ at our aeon's Big Bang also, providing agreement with the postulated \textit{Weyl curvature hypothesis} (\cite{Penrose2004}, subsection 28.8) giving our $2^{nd}$ Law in the form that we find it (with gravitational degrees of freedom enormously suppressed at the Big Bang). The vanishing of \textbf{\textit{C}} at ${\cal X}$  arises from the difference between the conformal weights of two proportional 4-valent tensor quantities \textbf{\textit{C}} and \textbf{\textit{K}}, where \textbf{\textit{K}} measures the ``graviton field" (analogously to the 2-valent Maxwell \textbf{\textit{F}} measuring the ``photon field"). Whereas \textbf{\textit{C}} directly measures conformal curvature, it is \textbf{\textit{K}} that satisfies a conformally invariant wave equation and thus has \textit{finite} values at ${\cal X}$ describing the ultimate gravitational radiation field at ${\cal X}$. 
We require $\mathit{\hat{K}} = \mathit{\hat{C}}$ with respect to the $\hat g$ metric, but \textbf{\textit{K}}=$\Omega$\textbf{\textit{C}} with respect to $g$, and it follows (from 
$\Omega=\infty$ at ${\cal X}$) that indeed \textsl{\textbf{C}}=0 at ${\cal X}$. Continuing into our aeon we find $\check{K}=\omega\check{C}$, which vanishes to second order, whence the gravitational degrees of freedom are greatly suppressed in the initial stages of our aeon - in agreement with the Weyl curvature hypothesis, yielding our $2^{nd}$ Law.

Nevertheless, despite \textsl{\textbf{C}}=0 at ${\cal X}$, there is information in the \textit{normal derivative} $\dot{C}$ of \textit{\textbf{C}} at ${\cal X}$ (being essentially \textit{\textbf{K}}, at ${\cal X}$), \textit{not} usually vanishing at ${\cal X}$. This enables gravitational information to pass from aeon to aeon. The \textit{magnetic part} of $\dot{C}$ at ${\cal X}$ determines the \textit{conformal curvature} of the 3-surface ${\cal X}$ (via the Cotton tensor, measuring ${\cal X}$'s departure of from conformal flatness), while the \textit{electric part} of $\dot{C}$ gives a direct measure of initial non-uniformity in the new dark matter, described by $\Omega$, this receiving an impulsive ``kick" in the direction of the gravitational impulse from the previous aeon (Fig.1). The field $\Omega$, initially taken as massless, has to acquire mass in the early stages of its evolution, by virtue of the equations (Appendix A and \cite{Penrose2010}, subsection B11).

\begin{figure}[htbp]
  \centering
  \includegraphics[width=120mm]{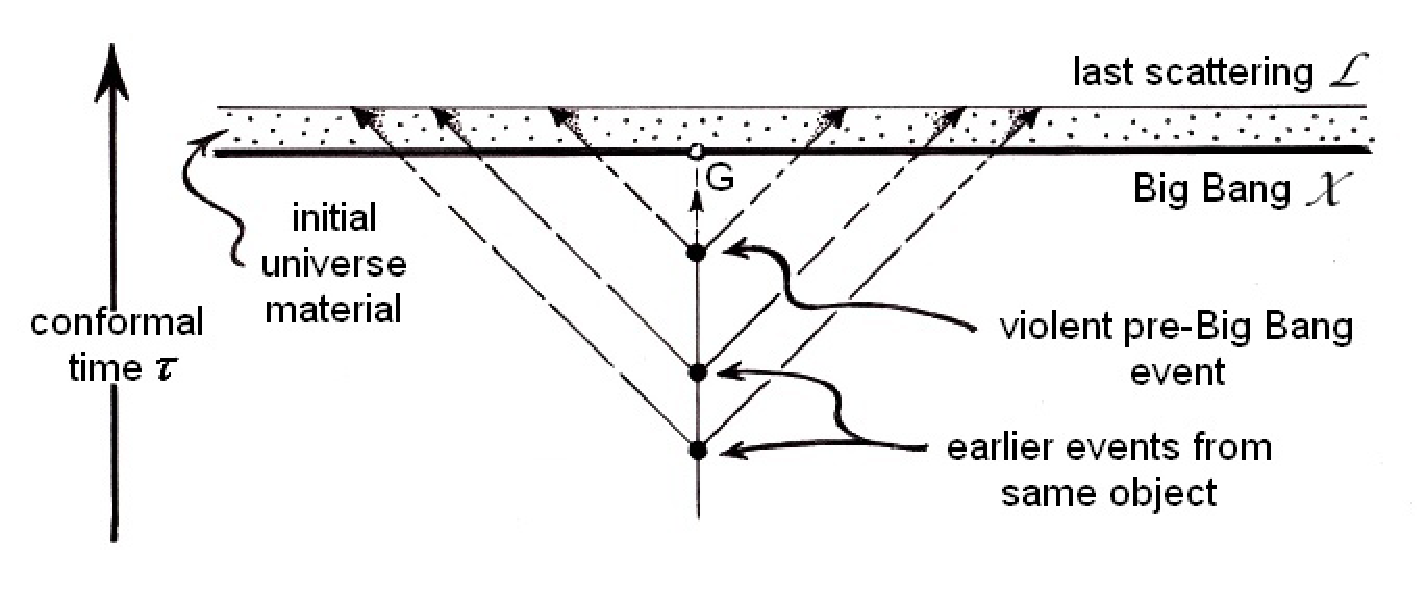}
  \label{fig:1}
  \caption{Conformal diagram (without inflation) according to CCC, where a pre-Big-Bang object (a galactic
  cluster with supermassive black-holes) is the source of three violent events.}
\end{figure}

This impulse in the initial material conveys considerable energy-momentum\footnote{See \cite{Penrose2011} for a discussion of awkward energy-momentum issues in $\Lambda>0$ cosmology.} along the outward null direction at ${\cal X}$, away from the source event in the earlier aeon (Fig. 1), reaching the last-scattering surface ${\cal L}$ as a largely isotropic outward-moving spherical shell of dark matter. The initial motion of this impulse, just to the future of ${\cal X}$, would be lightlike according to CCC, but would rapidly become dispersed by viscosity effects arising along with the appearance of \textit{rest-mass} in the $\Omega$-field in very early stages after crossover (presumably becoming significant at around the time when the new aeon cools to its ``Higgs temperature"). Details of this are not yet resolved in CCC theory, but something of this nature is to be anticipated. The general expectation is that by the time the last-scattering 3-surface ${\cal L}$ is reached, the impulse would be reduced to a small component in dark-matter motion, in the direction of the initial impulse, and that primordial acoustic damping would also have reduced its sharpness.

This expanding surface could be observed where it meets the past light cone of our present space-time location, seen as a circle that might be detectable in our CMB sky. This circle is the intersection of two spheres in the 3-surface $\cal L$, one of which describes where the future light cone of the violent pre-Big Bang event meets ${\cal L}$ and the other, where our own past light cone meets $\cal L$. The spreading spherical 2-surface itself would likely be locally much more energetic than other nearby variations in temperature in the early Big Bang, except at various places where it intersects other such spheres. Accordingly, this outward (almost impulsive) burst would normally have a rather closely uniform intensity over the whole outward-moving sphere in the initial material, except at these intersection places, though somewhat smeared out in its passage from Big Bang to last scattering. It would be seen as scattered light emitted preferentially in the outward direction, raising the CMB temperature somewhat, from our perspective, when there is a significant component of this impulse in our direction - the case for extremely distant sources. But when this motion is roughly in the opposite (i.e. outward) direction we would see a cooling of the CMB temperature - the case for relatively near sources. In each case, the raising or lowering of temperature is effectively a Doppler shifting resulting from the motion. Detailed estimates of the raising or lowering of the observed temperature from such a pulse would require a better understanding of the specific physical processes involved, these not having been fully investigated as yet. In broad terms, the expected general homogeneity of the pulse would lead to a narrow circular region in the CMB sky, which has lower temperature variance than normal. Appendix B explains the mathematical reason for the lowering of the temperature variance over such a CMB ring.

Of key diagnostic relevance here is the further point that such events ought to repeat themselves several times, if CCC is correct, with the circle's centre remaining at almost exactly the same point in the CMB sky for each burst. This is to be expected because such black-hole encounters would be likely to occur several times in the entire history of a single supermassive black hole. This would be especially the case for a black hole in a very large cluster of galaxies where there are numerous other supermassive holes available for such encounters with it. Moreover, with many galaxies within the same cluster, there could be several distinct supermassive black holes separately indulging in such activities within the same cluster. An entire cluster, if remaining bound in its remote future, would converge on a single point \textbf{G} of ${\cal X}$(Fig.1) because proper motions would normally be very tiny in comparison with the speed of light, so we can take the world-lines of all the relevant bodies (including black holes) to be essentially vertical and coincident in the conformal diagram. The point \textbf{G} would line up with a single point in our CMB sky and \textit{\textit{rotational symmetry}} about this point would extend to all the several gravitational wave bursts within the (bound) cluster, leading to a family of \textit{concentric} circles as the expected observational picture, for all these processes occurring within one galactic cluster. We might expect in some cases, perhaps from an eventually chaotic gravitational dynamics (or ultimate evaporation of mass, another feature of CCC; see Appendix A), that the galactic cluster might instead end up as several distinct ultimately bound portions separating from each other according to the exponential expansion of the later phases of the previous aeon. In such situations, the different portions, if each remains bound, would converge on separate but very close points of ${\cal X}$, and we might find that there are closely separated centres of several associated circular rings.

\begin{figure}[htbp]
  \centering
  \includegraphics[width=120mm]{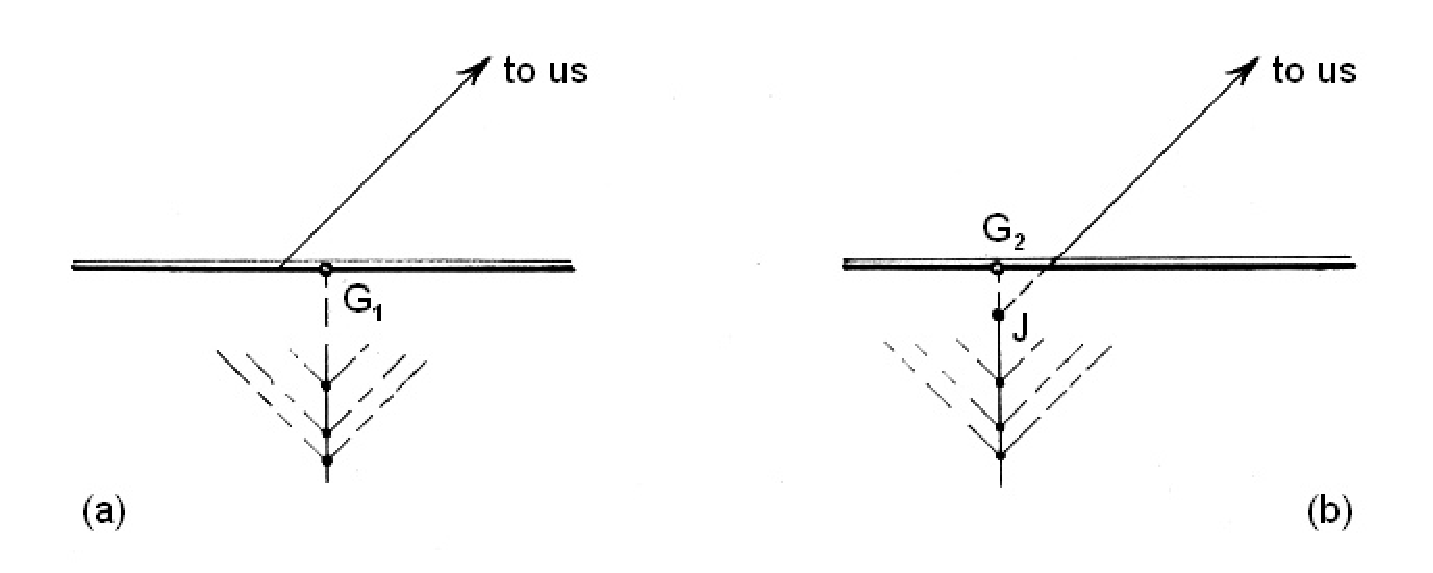}
  \label{fig:2}
  \caption{Conformal diagrams illustrating (a) a cooler-appearing source, the cluster's world-line lying entirely within our past light cone and not directly seen, and (b) a warmer-appearing source, the cluster's world-line intercepting our (extended) past light cone, so photons from the source might in principle be visible.}
\end{figure}

The point {\bf G} itself would \textit{not} normally lie on our past light cone, so it would only rarely be directly seen in these processes. The concentric rings are the things that CCC predicts to be observed, frequently widely spaced out from the central point, with nothing of particular observational significance normally to be seen at their centres, as arising from these predicted effects. However, we might anticipate, on the basis of CCC, that in some circumstances the central point \textit{could} exhibit effects of a quite different nature, coming from \textit{photon} emission from the galactic cluster in question. In Fig. 2, we have illustrated two different situations. In the case shown in Fig. 2a where \textbf{G} lies inside our past light cone, indicated by the point \textbf{G}$_1$, the entire history of the galactic cluster in question lies within our cone and nothing of relevance is to be seen at the central point; yet the impulsive gravitational events referred to above can still be seen in the case illustrated, as their light cones happen to intersect ours within $L$. In this case we would expect the rings to be cooler than average. But in the case Fig. 2b, where \textbf{G} lies \textit{outside} our past light cone, indicated as \textbf{G}$_2$, taken to be near enough that the impulsive events can still be seen, we find that the extension of our past light cone into the previous aeon \textit{does} encounter the world-line of the galactic cluster (point \textbf{J}), so that photon emissions from that cluster could in principle be observable to us. Such emissions could be ordinary starlight, supernova explosions, or active galactic nuclei, but most interestingly, we might conceivably observe the extremely late history of the galactic cluster in question through \textit{Hawking radiation} from its supermassive black holes! Although such radiation is of such an extraordinarily low temperature that its direct detection would normally be taken as an utter absurdity, we must bear in mind that virtually the entire mass-energy of such a hole, perhaps of up to $10^{10}$ solar masses or more, is eventually deposited in such low-energy photons. Since the entire late history of such holes (which might last for up to, say, some $10^{100}$ years) could be compressed into a very tiny visual region, from our vantage point, the observational status of this process is something to be considered. Any of the aforementioned possibilities for photon emission might have to be taken into consideration, especially if many galactic clusters occur together within some small visual angle. It should be borne in mind that such radiation could only be in principle visible to us when our past light cone intersects the world line of a galactic cluster, which is the second case considered above, $G=G_2$, and the central point \textbf{J} in Fig. 2b is what would be observed. This is the case for especially distant sources where the circular rings would be expected to be of \textit{above-average} temperature. We propose to explore this intriguing observational issue in future investigations.

On the whole, it appears that on the basis of CCC the major small-scale disturbances of initial material that we observe in the CMB would come from back-hole encounter processes of the type described here, although irregularities in the overall matter distribution and electromagnetic effects ought also to contribute. If we assume the circular rings to be the major small-scale effect, we have many independent overlapping families of circles of low temperature variance, and varying average temperatures, somewhat resembling a pond after a period of rain, but where the sources of the raindrops would be collected, to a significant extent, at individual points, providing the tendency to concentricity referred to above. This concentricity is an implicit feature of the claimed predictions of CCC \cite{Penrose2010}, but not spelled out until in 
\cite{Gurzadyan2010a}. In Section 2, we point to clear evidence of numerous concentric low-variance circles in the WMAP data whose prevalence depends upon their being actually \textit{circular}, and that the distribution of these concentric sets is manifestly non-Gaussian.

It would be anticipated from CCC, also, that the mean temperatures of the different circles in a concentric set should be somewhat correlated with each other, as also would be the mean temperatures resulting from events that occur in the same general space-time region in the previous aeon - if it may be assumed that these mean temperatures are indeed due largely to the angle between the direction of the impulse in the initial dark matter and our line of sight, i.e. to the angle of intersection between the spheres in ${\cal L}$ representing the impulse and our past light cone. Curiously, as remarked above, it is the more distant sources that would be generally seen having the higher temperature, since for them the impulse is directed more towards us, and correspondingly, the very closest ones should appear the coolest. There will be no dimming due simply to the source's distance, as can be seen from the geometry of the conformal diagram. Moreover, so long as the spatial geometry can be considered Euclidean, we expect a somewhat greater number of distant than close ones - and therefore more warmer than cooler ones - for a given circle size, again owing to this geometry.  There may be other factors influencing this temperature, but nevertheless this behaviour of the mean temperatures over the low-variance circles is also of diagnostic value, and we examine such matters in Section 2. In fact, the observations do appear to be in general agreement with CCC's expectations. 

Some possible complicating factors need to be mentioned here. One is that physical processes between the Big Bang ${\cal X}$ and the last scattering surface $\cal L$, such as acoustic waves and damping, would lead to a smearing out of the initially impulsive circular signal, up to the scale of the order of the horizon size at ${\cal L}$ i.e. at about $1^\circ$ in the CMB sky. Such acoustic effects would allow a refinement in our analysis, but certainly do not invalidate it, as our rings spread over 
$0.5^\circ$ in the radial dimension, and therefore average over that scale. Such refinements could be a basis for future work. Another consideration is that, at larger scales, matter-density irregularities in our own aeon might perhaps lead to significant distortions of circular patterns at $\cal L$, owing to the (astigmatic) gravitational lensing effects (e.g. \cite{Gurzadyan2007}). The largest known sources of such density irregularities might occur in connection with the observed large voids \cite{Gurzadyan2009a}\cite{Gurzadyan2009b}, see also \cite{Cap}. There could also be lensing of a similar kind arising in the aeon prior to ours but, for various reasons and unknowns, such lensing would be harder to account for. Lensing effects would generally be expected to be to be too small to be of great relevance to the WMAP observations considered in this paper, but their possibility should not be ignored. We return to this question in Section 2.

Effects of this nature in our own aeon ought in any case to be generally larger than those in the previous aeon, owing to the greater relevant light-ray ``lengths", in the conformal diagram in our own aeon. According to a recent estimate due to \cite{Tod2011} and by \cite{Nelson}, our present temporal location is about 3/4 of the way up our aeon's conformal diagram, so there might be about 3 times as much conformal ``length" for lensing in ours than the previous aeon, for a black-hole encounter taking place at around the corresponding time to ``now" in that aeon. For later-occurring such encounters, the ratio would be even greater. Similar considerations apply also to the maximum angular size of a low-variance circle produced by a means such as that described above. Assuming, as above, that the previous aeon was similar to ours and that the co-moving time lines more-or-less match from aeon to aeon, and that we are indeed now about 3/4 of the way up our aeon's conformal picture, black-hole collision events occurring no earlier than the corresponding stage of the previous aeon could not give rise to low-variance circles of an angular radius of more than about $20^\circ$ (\cite{Tod2011}, compare \cite{Nelson}). For later events in the earlier aeon, the limit would be smaller. We shall see in Section 2 that the observations are very consistent with this restriction.

\section{Concentric sets of low-variance circles in the CMB}

Fig. 3a shows the distribution over the sky (without the Galactic disc region) of concentric sets containing \textit{three or more} low-variance circles (see Section 0), the left-hand figure showing the location of the centres and the right-hand one the actual rings. The colour bar scale is in $\mu$K within $3\sigma$ range. WMAP's 7-year W band (94 GHz) data are used (monopole and dipole extracted), as the least contaminated by the Galactic synchrotron emission and having the highest angular resolution. We have tested the CMB temperature variance properties also for the WMAP's V-band (61 GHz) and for BOOMERanG maps (150 GHz) \cite{Gurzadyan2010a}, for maps smoothed by two procedures, via averaging over 8 neighbours and by the algorithm of \url{http://lambda.gsfc.nasa.gov/product/map/current/m_products.cfm}. The main results remain robust also when the KQ85 mask replaces ours. The role of noise was included via overlapping the difference maps (W1-W2) with simulations based on the empirical power spectrum. To test the role of noise, we have run the same procedure also for W1-W2 \textit{difference} maps: in that case Fig. 3a is blank, and so also is the map for depth over $10\mu$K, the first two points appearing at $6\mu$K. For the circles intersecting the masked region $|b|<20^\circ$ only the pixels outside that region (necessarily greater than semi-circular) are taken into account.

\begin{figure}[!htbp]
  \centering
  \includegraphics[width=140mm]{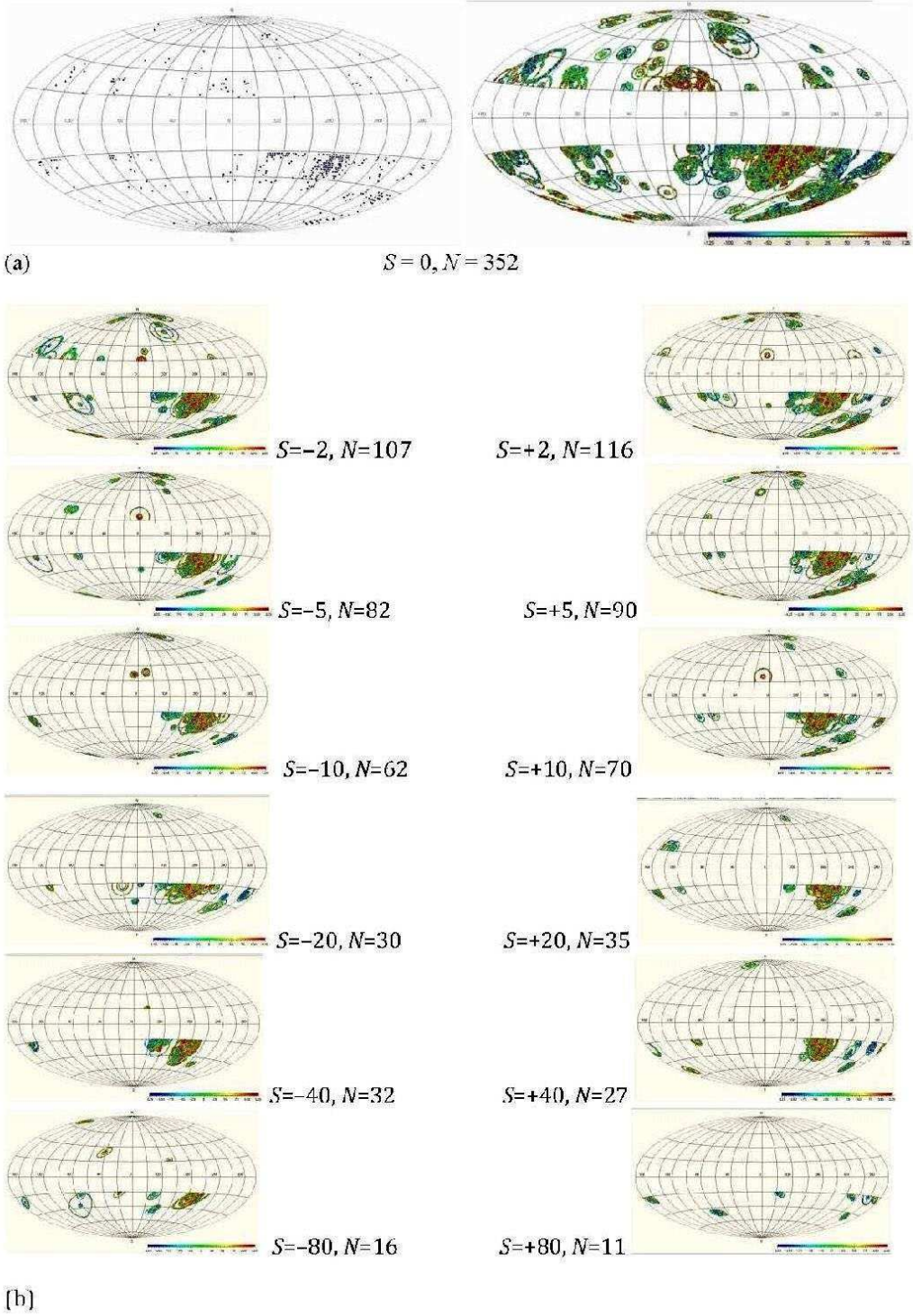}
  \label{fig:3}
  \caption{(a) The sky distribution of concentric sets containing three or more circles of variance depth over $15\mu$K: the left-hand figure indicates the positions of the 352 centres; the right-hand one exhibits the actual circles. (b) The same analysis applied to concentric elliptical shapes, obtained by sky-twisting with various shifts $|S|=2, 5, 10, 20, 40, 80$, showing a sharp fall-off in numbers $N$ of centres, with increasing shift.}
\end{figure}

A distinct inhomogeneity/clustering of the centres of the concentric sets - 352 in total - is visible in Fig.3a, which indicates their clearly \textit{non-random} distribution\footnote{Simulations that we have performed using the empirical power spectrum, beaming (\textit{synfast}, $R=0.02$, \url{http://lambda.gsfc. nasa.gov/toolbox/tb_camb_form.cfm}) and WMAP's noise, led, as expected, to random distributions of the centres of the concentric low-variance sets.}. This manifest inhomogeneity is, by itself, hard to accommodate within the conventional picture that the temperature fluctuations in the CMB arose from early quantum fluctuations in an inflaton field. For that, we would expect a distinctly homogeneous/isotropic distribution over the whole sky. Moreover, quite apart from the specific considerations of the present paper, it has long been noticed that there are hard to explain low-multipole deviations from the conventional picture's expectations (see, for example, \cite{Starkman}).

Such anisotropies, although not necessarily anticipated on the basis of CCC, are perfectly consistent with it. Vast regions in the previous aeon, containing many sources of sequential emissions of the type considered in Section 1 could result from great concentrations of particularly large galactic clusters, so that multiple encounters between supermassive black holes would be common there. It is interesting that an unexpectedly huge concentration of quasars-and therefore of supermassive black holes-has recently been discovered \cite{Clowes} in our current aeon, so it is perhaps not so surprising that the large concentration that we appear to be seeing here could have been present in the previous aeon. The sizes of galaxies and  clusters in that aeon would depend upon the nature of the small-scale disturbances that the initial dark matter encountered in that aeon's early stages and those disturbances would, according to CCC, come mainly from the effects of supermassive black-hole encounters in the aeon prior to that. It would be a complex matter to calculate how such effects propagate from aeon to aeon, but there is no particular expectation of overall uniformity. What CCC \textit{does} require is that the general character of each aeon should propagate from each to the next, but there is no \textit{a priori} reason to expect an entirely homogeneous large-scale spatial geometry or uniformity in galactic cluster sizes.

In order to show that the number and distribution of the 352 centres found in Fig. 3a are not spurious statistical effects without any astrophysical cause like those of CCC, we have adopted a novel \textit{sky-twist} procedure, aimed at testing whether the numerous centres of multiple low-variance rings depend upon their being \textit{circular} rather than of some other shape\footnote{Our sky-twist procedure provides a more realistic search than for, say, squares or triangles, showing that even for concentric sets of slightly deformed circular shapes, the numbers found drop off dramatically.}. This is a simple operation that translates our search for multiple concentric low-variance circular rings into a corresponding search for concentric low-variance elliptically shaped rings.

We use standard spherical polar angles $(\Theta,\phi)$ for the celestial sphere, here related to conventional galactic coordinates $(b,l)$ by $b=90-\Theta$, $l=\phi$ all angles being in degrees. The \textit{sky-twist} is a transformation of the celestial sphere, determined by a given angle $S$ referred to as the \textit{shift}, again in degrees, whereby the celestial point $(\Theta,\phi)$ is mapped to $(\Theta^\prime,\phi^\prime)$, according to
$$
\Theta^\prime = \Theta,  \qquad\qquad  \phi^\prime = \phi + S\Theta,
$$
with inverse transformation
$$
\Theta = \Theta^\prime,   \qquad\qquad  \phi = \phi^\prime - S\Theta^\prime,
$$
This is an area-preserving transformation of the CMB sky which simply rotates each latitudinal circle of points on the celestial sphere, given by some constant value of $\Theta$, through an angle $(1/90)S\Theta$. The particular set of 10885 points that we examine as potential centres of low-variance circles all lie on latitudinal circles where $\Theta$ is an integral multiple of 1.5 (as stated in Section 0), and these latitudinal circles simply rotate by $S\Theta$. The key feature of the sky-twist is that general circles on the celestial sphere are transformed to \textit{elliptical} shapes, and also conversely, so that when the algorithms that we have adopted for searching for concentric low-variance \textit{circular} rings in the CMB sky are applied to the twisted sky instead, this is, by the inverse transformation, completely equivalent to searching for such \textit{elliptical} shapes in the \textit{true} CMB sky. The inverse sky-twist reveals the pattern of these concentric low-variance elliptical shapes in the CMB.

It is clear that, for a given value of $S$, the elliptical shapes obtained in this way are just as numerous as exact circles on the celestial sphere, each being determined by its centre and a radial measure of size. Accordingly, if one is to believe that the number and distribution of concentric low-variance circular rings in Fig.3a are simply some kind of statistical artefact, then one would expect to find something similar for the \textit{elliptical} shaped rings obtained via the sky-twist procedure, for each value of $S$. However, the results obtained show a dramatic difference, for various values of $S$ (namely $S=\pm2$, $\pm5$, $\pm10$, $\pm2$0, $\pm40$, and $\pm80$), these being depicted in Fig.3b. We note that even for a shift as small as $S=\pm2$, (for which the departure from circularity is around one or two per cent), the number of centres of $\geq3$ low-variance rings, drops to less than a third of the number (352) found for exact circles. With increasing ellipticity (i.e. increasing $|S|$) the numbers continue to drop away consistently.

We should stress that although our sky-twist procedure does involve circle searches in modified CMB skies, we are not using these twisted skies as ``simulated skies" in the manner of conventional tests for statistical significance of CMB features. The twisted skies may well have significantly differing power spectra from that of the true CMB, but this is not relevant here, because the sky-twist procedure is a search for \textit{elliptical} shapes in the true sky, which simply happens to be equivalent to circle searches in various twisted skies.

Although we refer to the distorted circular rings obtained by the sky-twist as ``elliptical shapes", they are not strictly ``ellipses", as exact non-circular ellipses are not found on spheres. However the sky-twist does give exact \textit{infinitesimal} ellipses. Simple geometrical considerations allow us to calculate the eccentricities of these small ellipses. We find, defining
$$
s = \frac{1}{180} |S| \sin \Theta^\circ,
$$
that the ratio $\rho$ of minor to major axis of the small ellipse (the \textit{eccentricity} being $\sqrt{1-\rho^2}$) is
$$
\rho = (\sqrt{1+s^2}-s)^2.
$$
Taking the main ranges of $\Theta$ of interest for Fig.3b to be given by $20\leq |b| \leq 60$, i.e. $30\leq \Theta \leq 70$ or $110\leq\Theta\leq150$, we display the values of $\rho$ at the limits of those values in Table 1. 

\begin{table}[!htb]
\begin{center}
\begin{tabular}{l|cc}
             &	$\Theta =70$ or 110	& $\Theta=30$ or 150\\
\hline
\hline
$S = \pm2$	 &     0.979	          &    0.989\\

$S = \pm5$	 &     0.949	          &    0.973\\
$S = \pm10$	 &     0.901	          &    0.946\\
$S = \pm20$	 &     0.812	          &    0.895\\
$S = \pm40$	 &     0.661	          &    0.801\\
$S = \pm80$	 &     0.444	          &    0.643\\
\hline
\end{tabular}
\end{center}
\caption{For various shifts $S$, values of $\rho$ (minor to major axis ratio) for small ellipses relevant to Figs. 3 and 4 are listed.}
\end{table}
We clearly see from Fig 3b that for increasing ellipticity (i.e. decreasing $\rho$) the number of centres of $\geq3$ concentric low-variance rings falls away sharply. When $S=\pm2$, $\pm5$, or $\pm10$, deviations from circularity are less than 10\%, and such relatively small non-circularities are not obvious to the eye. Yet with $S=\pm10$ the number of centres has dropped by more than a factor of 5, indicating a distinct signal of support for the hypothesis that the concentric \textit{circular} rings of Fig.3a result from genuine astrophysical/cosmological process, rather than being statistical artefacts.

\begin{figure}[!htbp]
  \centering
  \includegraphics[width=140mm]{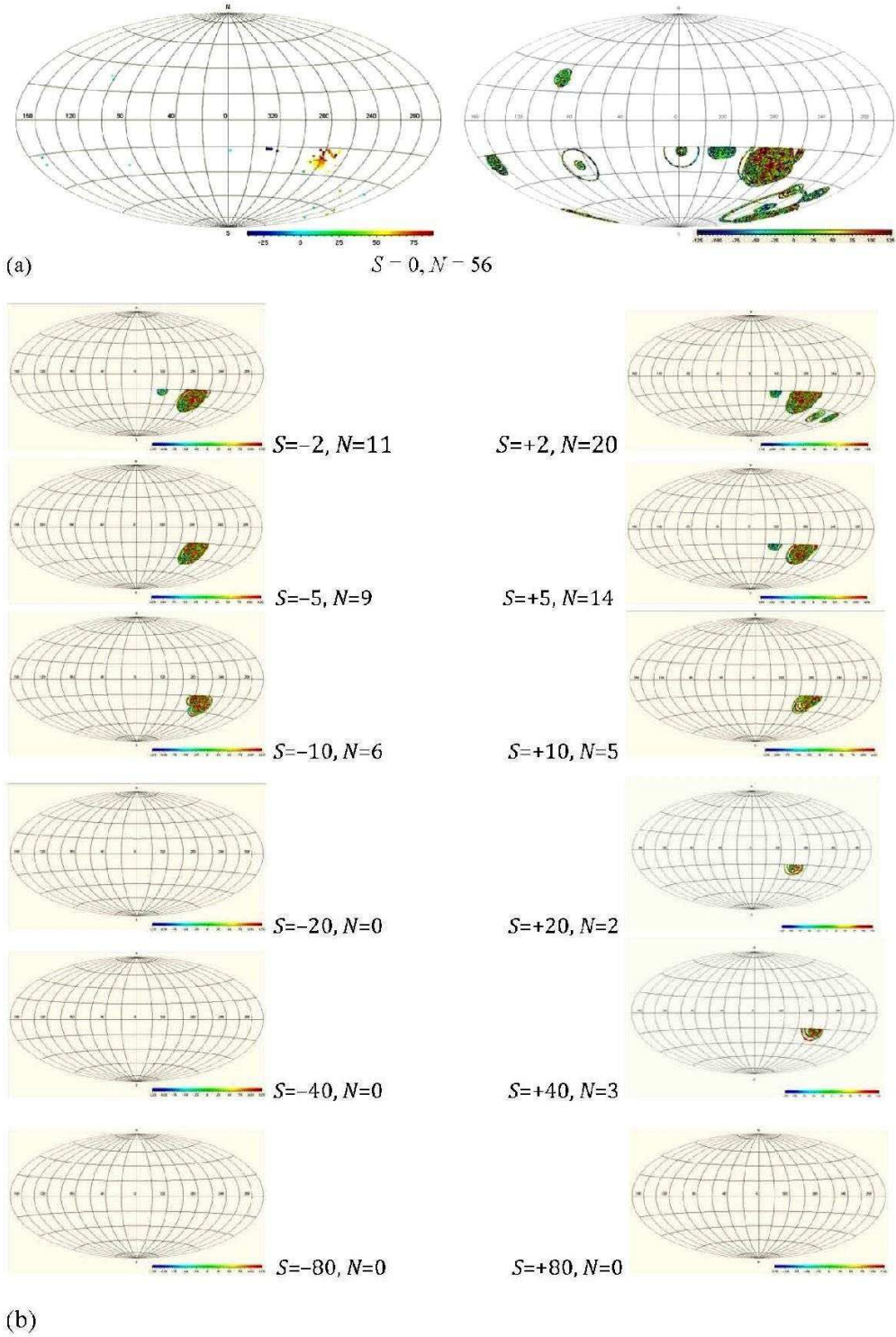}
  \label{fig:4}
  \caption{The same as in Fig.3 but for the concentric sets containing four or more circles of variance depth over $15\mu$K. }
\end{figure}

Nevertheless, the clumping that we see clearly in Fig.3a, does not entirely disappear, even at $S=-80$. As a possible explanation, gravitational lensing should not be entirely ruled out, but it is hard to see how this could achieve a magnitude necessary to distort CCC's circular rings to the observed elliptical ones. More probable - at least for the most crowded area of centres of low-variance triples of circles on the right just below the excluded region (henceforth referred to as \textit{region X}, concentrated around the location $l=280$, $b=-30$) - is that large numbers of closely concentric low-variance circles might conspire, through the close overlapping of adjacent low-variance circular rings, to give rise to occasional spurious triples of concentric low-variance elliptical shapes. It is not our purpose, in this paper, to try to provide CCC-explanations for all observed effects, but we point out that there are genuine effects - at least partially anticipated by CCC - that need explaining, whether by CCC or by some other type of process. 

The case for the genuine nature of these concentric rings appears to be even stronger when we examine the centres of \textit{four} or more low-variance circles, as seen in Fig.4a, where we still find 56 such centres. The inhomogeneity is even more striking here than in Fig.3a and it is yet harder to see how the patterns arising could result from a random initial input, such as the quantum fluctuations of inflation. According to CCC, on the other hand, such an inhomogeneous pattern could result from distinctively non-uniform galactic matter distributions in the previous aeon, black-hole encounters occurring more frequently or with greater intensity in some regions than others. On this basis, there appears to have been a very large previous-aeon source concentration in the direction of region \textit{X}. Moreover, these sources would have been particularly distant, in accordance with our considerations near the end of Section 1, owing to the rings' higher temperatures indicated by the largely red or yellow colour-coding for the centres in region \textit{X}. There are also other (somewhat less prominent) regions of the CMB sky of a similar character. On the other hand, there are different regions, such as occurs just below the galactic disc, just to the right of centre (henceforth referred to as \textit{region Y}, concentrated around $l=330$, $b=-20$), where we find a deep blue colour, coding cooler temperatures, indicating, according to CCC, a relatively nearby smaller collection of previous-aeon sources.

Again, we demonstrate the implausibility of these features simply being some kind of statistical artefact by comparing the number of centres of quadruple circular low-variance rings (56) with the corresponding numbers of concentric elliptically shaped low-variance rings (Fig.4b). We find a drop-off with shift rather sharper, even, than with the triple rings. We note that the centres disappear completely when $S=\pm80$ which, according to Table 1, refers to a search for quadruple concentric elliptical shapes, each about twice as long as it is wide ($\rho\approx\frac{1}{2}$). This happens also with $S=-20$ and $S=-40$. When $S=+20$ we do find two surviving centres and three when $S=+40$, located in different parts of region {\it X}. The rings in these two pictures are probably not significant, their centres lying very close to the excluded region, so only about half of each ring is examined, the role of elliptical distortion being much reduced. In all these considerations is the possibility of complication due to gravitational lensing, such as a super-dense closer region {\it Y} influencing distant signals from {\it X}, but such effects should be very small, not much affecting our conclusions. In any case, CCC does appear to provide a consistent picture, though with a surprising inhomogeneity in previous-aeon galactic clustering.

It should be noted that the crowding of low-variance rings around region $X$ cannot be explained simply by a suggestion that this area is of low variance as a whole (for some reason not connected with CCC), so that triples and quadruples of concentric low-variance rings might arise fortuitously. In fact, this area is of slightly \textit{higher} variance than that in the CMB sky generally. If we take that part of it lying in discs of successively increasing angular radius, each centered at ($280.00^\circ, -30.00^\circ$) close to the middle of region $X$, we find respective variances
$$
3^\circ\colon \; 218.95𝜇\mu\mathrm{K}, \qquad 4^\circ\colon \; 178.13𝜇\mu\mathrm{K}, \qquad 5^\circ\colon \; 154.93𝜇\mu\mathrm{K}, \qquad 6^\circ\colon \;  138.50𝜇\mu\mathrm{K}.
$$
This may be compared with the sigma of the CMB sky, which is around $110𝜇\mu$K (the sigma of the circles we examine being, on average, less than about $95𝜇\mu$K).

In Section 1, we noted CCC's prediction of more warm than cool low-variance circles in the CMB. Indeed, we find that the mean temperature excess, over the low-variance sets of 3 or more rings in WMAP data is $9.8 \mu$K, which is higher than that for those in 20 simulations (see footnote 6) where we find an average of $1.8 \mu$K. We also note the important point that whereas many low-variance circular rings of radius $>12^\circ$ are seen in the WAMP data, \textit{none} have angular radius $>15^\circ$. This is consistent with CCC's expected cut-off at around $20^\circ$ as remarked in Section 1, but difficult to square with inflation \cite{Tod2011,Nelson}). 

The region $X$ happens to be close to the location (excluded by the KQ85 mask) of the Large Magellanic Cloud (LMC, $l=280$, $b=-33$). However, this association with LMC is certainly fortuitous, the rings themselves (and most of their centres) lying almost entirely outside this location. Such a putative physical association would require sources of extremely violent \textit{sequential} explosions, far exceeding the effects of supernovae, whereas no evidence of such repeated explosive events, or of the intergalactic dust needed to make them visible, is seen either in association with LMC or any other galaxy.  Moreover, neither the Small Magellanic Cloud, the Andromeda galaxy nor the Triangulum galaxy (of our Local group) show any sign of being associated with centres of multiple low-variance rings. Furthermore, the observed proper motion of LMC of about 500km/s would lead to about 10 arcmin parallax for the circle centres, destroying concentricity of the rings.

Region $X$ raises another issue, however, concerning the possibility of previous-aeon photon emissions (unrelated to LMC). Since the concentric rings of region $X$ are \textit{warmer} than average (left-hand Fig.4a), CCC provides us with the prospect of sources of unusual radiation within the area of $X$, coming directly from very late photon emissions from the numerous large previous-aeon galactic clusters, as indicated in Fig. 2b. Curiously, this galactic conglomeration would have to be beyond the ``particle horizon" of conventional cosmology, as indicated by the point $G_2$ of Fig.2b. This intriguing issue is well deserving of further study.

Possible associations between, the low-multipole temperature anomalies pointed out by others \cite{Starkman} - e.g. the ``quadrupole plus octopole anisotropy" - and the anisotropies found in Figs. 3a and 4a are worthy of further exploration. CCC's interpretation is that the warm region $X$ comes from a huge, very distant previous previous-aeon galactic conglomeration, and that the cool region $Y$ corresponds a smaller, much closer one.

\section*{Appendix A: CCC mathematical details}

Some, but not quite all, of the relevant mathematical equations underlying the predictions of CCC are to be found in the Appendices to \cite{Penrose2010}. The main issues of relevance are provided here. As stated in Section 1, CCC posits that a conformal space-time ``crossover 3-surface" ${\cal X}$ smoothly joins the open space-time region $\check \Re$ (with metric tensor $\check g_{ab}$), immediately following the Big Bang of our aeon, to another open region $\hat \Re$ containing the remote future of an aeon (having metric tensor $\hat g_{ab}$) just prior to ours. An open 4-region $\Re$ (a ``collar") with metric tensor $g_{ab}$, contains ${\cal X}$. Indices are raised or lowered using the metric whose ``hatted" nature agrees with that of the kernel symbol (e.g. ${\check{R_{abc}}}^d=\check g^{de}\check{R}_{abce}$); sign conventions, etc. are as in \cite{Penrose1984}. Where the regions of definition of two of the metrics match, they are conformally related according to
$$
\hat g_{ab} = \Omega^2 g_{ab} \quad \mathrm{and} \quad \check g_{ab} = \omega^2 g_{ab}.
$$
We have $\Omega\rightarrow\infty$ and $\omega\rightarrow 0$ as ${\cal X}$𝒳is approached and we impose the \textit{reciprocal hypothesis}:
$$
\Omega\omega = -1.
$$
This extends the definitions of $\hat g_{ab}$ and $\check g_{ab}$ to the entire $\Re$, except that $\Omega$ and $\hat g_{ab}$ are infinite at ${\cal X}$, carried across by the demand that $\omega$ be smooth with non-vanishing gradient at ${\cal X}$𝒳(where $\omega = 0$), and we note that both $\Omega$ and $\omega$ change sign when ${\cal X}$ is crossed, whence the minus sign in the relation between them. Of significance is the 1-form
$$
\mathrm{\Pi} = \frac{d\Omega}{\Omega^2-1}𝚷= \frac{d\omega}{1-\omega^2}, \quad\quad \mathrm{i.e.} \quad \mathrm{\Pi}_a = \frac{\nabla_a\Omega}{\Omega^2-1} = \frac{\nabla_a\omega}{1-\omega^2},
$$
which is smooth across ${\cal X}$ and unaltered under interchange of $\Omega$ with $-\omega$.

We require a \textit{positive} cosmological constant $\Lambda$ in both aeons, adopting for convenience \textit{cosmological units} (so Newton's constant $G$ could vary, if need be) where
$$
c = 1, \qquad \hbar = 1, \qquad \Lambda = 3.
$$
The \textit{spacelike} nature of ${\cal X}$ is thereby ensured (\cite{Penrose1984}, p.353); moreover $\nabla^a\omega$, and therefore also $\mathrm{\Pi}^a$, is the \textit{unit normal} to ${\cal X}$ at all points of ${\cal X}$: 
$$
\nabla^a\omega\nabla_a\omega =𝒳1 \; \mathrm{at}\; {\cal X}, \qquad \mathrm{i.e.} \; \mathrm{\Pi}^a\mathrm{\Pi}_a = 1 \; \mathrm{at} \; {\cal X}.
$$

In fact, we need this to second order at $\cal X$, as part of the requirement for uniqueness of propagation across $\cal X$ (compare \cite{Penrose2010}, p.246); more specifically: 
$$
\mathrm{\Pi}^a\mathrm{\Pi}_a𝒳= 1 + Q\omega^2 + O(\omega^3) \; \mathrm{at} \; {\cal X}, \qquad \mathrm{i.e.} \; \nabla^a\omega\nabla_a\omega = 1 + (Q - 2)\omega^2 + O(\omega^3)  \; \mathrm{at} \; {\cal X}
$$
where $Q$ is a given (positive) universal constant\footnote{The necessity of having some non-zero quantity $Q$ in this relation was pointed out by K.P. Tod. The constancy of this $Q$ over $\cal X$𝒳 is an assumption which might turn out to have to be generalized. The postulated ``anti-Higgs" mechanism in the very remote future of the previous aeon is a related issue.
}, taking some appropriate value for agreement with the details of rest-mass re-emergence in the early universe, presumably in accordance with a suitable Higgs-type mechanism. The role of this condition - the \textit{suppressed rest-mass hypothesis} - will appear later. We remark that $\omega$ is a suitable ``conformal time coordinate" in $\Re$, with $\omega = 0$ at $\cal X$ (see Fig.1, with the choice $\tau = \omega$).

The mathematical formulae describing crossover between aeons, determining the equations of state for the emergent material within $\check \Re$, are governed by simple principles, though resulting in expressions of moderate complexity. We keep in mind that rest-mass ought to be absent at $\cal X$ in order that the philosophical basis of CCC be maintained, this demanding that local clocks become physically impossible at $\cal X$, and the conventional meaning of a ``physical metric" at $\cal X$ evaporates, leaving us with the \textit{conformal} structure that allows smooth transition from aeon to aeon. This vanishing of rest-mass at $\cal X$ is taken in an \textit{asymptotic} sense, so rest-mass tends to zero suitably fast as $\cal X$ is approached. We find that the equations force rest-mass to \textit{reappear} following $\cal X$, even if there is assumed to be exactly no rest-mass within $\hat \Re$. However, if we adopt the above ``suppressed rest-mass hypothesis", the emergence of rest-mess in $\check \Re$ may be effectively delayed, for suitable $Q$, until the ``Higgs time" of current particle-physics ideas\footnote{The details of this, in relation to the choice of $Q$ are not settled, as yet, depending on various unknown factors, especially concerning the emergence of rest-mass in the initial form of dark matter. This also relates to the change-over from radiation dominance to matter dominance in the early universe, a curious point of interest being that this appears to occur at a time of general order unity, in cosmological units.}.

In his proposal \cite{Tod2003}, Tod revealed severe consistency restrictions on the equations of state following the Big Bang if it is assumed (as CCC actually demands) that the Weyl curvature is exactly zero at $\cal X$. Accordingly, CCC adopts a specific scheme that ensures consistency of the particular equations of state that ensue. This is that those equations (Einstein's equations with $\Lambda>0$ and massless sources) that apply in $\hat\Re$ with its $\hat g_{ab}$ metric, i.e. within the remote future of the previous aeon, would \textit{continue} into the early part of $\check\Re$ if the prior $\hat g_{ab}$ metric is continued into $\check\Re$. This continuation also propagates the specific \textit{solutions} to these equations that apply within $\hat\Re$, propagating uniquely into $\check\Re$ according to these $\hat g_{ab}$-metric equations. In order for propagation across $\cal X$𝒳to be achieved smoothly, despite the infinite character of $\hat g_{ab}$ ($\Omega = 0$), the equations are translated into $g_{ab}$-metric form to cover this transition. When re-interpreted in $\check\Re$ we obtain a $\hat g_{ab}$-picture of a universe collapsing inwards from infinity, resembling the \textit{time-reverse} of a remote future of an exponentially expanding previous aeon (with its $\hat g_{ab}$-metric description of $\hat\Re$), but with one key difference: its effective gravitational constant has become \textit{negative} (basically, owing to the negative nature of the $\Omega$-field in $\check\Re$). Initially - so long as rest-mass can be maintained at zero - a physical interpretation is available \textit{whichever} metric $\hat g_{ab}$ or $\check g_{ab}$ is adopted within $\check\Re$, albeit very different in appearance. While the $\hat g_{ab}$-picture of $\check\Re$ is a universe collapsing inwards from infinity, in $\check g_{ab}$ it expands out from a singularity. The viewpoint of CCC is to choose the $\check g_{ab}$ picture within $\check\Re$, in order to continue this evolution in a physically appropriate way, the effective gravitational constant becoming \textit{positive} after ${\cal X}$ is crossed.

The reciprocal hypothesis can enable this re-interpretation to be achieved uniquely (the new conformal factor $\omega = -\Omega^{-1}$ allowing us to translate uniquely from $\hat g_{ab}$ to $\check g_{ab}$ via $g_{ab}$) if enough conditions can be imposed on the freedom of choice of $\Omega$ to make \textit{this} unique. This is partially achieved by first requiring the intermediate $g_{ab}$ metric to have a scalar curvature $R$ that satisfies the same relation $R = 4\Lambda$, or
$$
\frac{1}{12}\;R = 1
$$
(in the cosmological units adopted here), implied by an imposition of the $g_{ab}$-metric Einstein equations with massless sources and cosmological constant $\Lambda(=3)$. Although one of the assumptions of CCC is indeed a gradual decay of rest-mass in the extremely remote future of each aeon\footnote{This ``anti-Higgs mechanism" is an asymptotic decaying away of rest-mass itself, not a decay of massive particles into massless ones. Though neither supported nor refuted by observational evidence, this proposal perhaps gains some justification from the fact that, when a cosmological constant is incorporated into particle physics, the local dynamical symmetry group is more properly the de Sitter rather than Poincare group, rest-mass not being a Casimir operator of the de Sitter group, and so not automatically constant for a stable particle.}, CCC yet lacks a specific theory for the time scale of this process.

Provisionally, we avoid the issue by assuming that the region $\hat\Re$ is chosen where no massive particles happen to be present - which may be considered to be at \textit{all} places near enough to ${\cal X}$ if we are satisfied with regarding this masslessness as an \textit{asymptotic approximation}, where the complications arising from the actual presence of rest-mass, decaying away asymptotically to become zero at ${\cal X}$, will not substantially alter the conclusions. This allows us to take the physical energy tensor $\hat T_{ab}$ to be trace-free in $\hat\Re$:
$$
{\hat T_a^a} = 0.
$$

The equation satisfied by $\Omega$ expressing the requirement $\Lambda = 3$ in the $g_{ab}$-metric is the second-order partial differential equation referred to in \cite{Penrose2011} as the ``$\varpi$-equation", an instance of the \textit{Yamabe equation} (\cite{Yamabe}) of differential geometry, having \textit{physical} interest as the equation for a conformally invariant (under rescaling $\tilde\varpi = \tilde\Omega^{-1}$ for metric rescaling $\tilde g_{ab} = \tilde\Omega^2g_{ab}$) self-coupled scalar field $\varpi$:
$$
(\square + 2)\varpi = 2\varpi^3,
$$
with $\square=\nabla^a\nabla_a$. It may be remarked that when $\Omega(=\varpi)$ indeed satisfies this equation, then we can recover $\Omega$ from the 1-form $\mathrm{\Pi}$ 𝚷by means of the formula
$$
2\Omega = \frac{\nabla^a\mathrm{\Pi}_a}{1-\mathrm{\Pi}_b\mathrm{\Pi}^b}.
$$

We recall, from the suppressed rest-mass hypothesis that the denominator vanishes to second order in $\omega$. By the reciprocal hypothesis, $\Omega$ has just a \textit{simple} pole in $\omega$, so the divergence $\nabla^a\mathrm{\Pi}_a$ of $\mathrm{\Pi}$must vanish to first order; specifically:
$$
\nabla^a\mathrm{\Pi}_a = 2Q\omega + O(\omega^2),
$$
so we find that the above displayed expression for $2\Omega$ is a version of ``0/0" near ${\cal X}$, so it is a somewhat subtle matter that $\Omega$ is retrieved from this expression rather than $-\omega$. This is to be expected since $\Omega$ and $-\omega$ are on an equal footing with regard to $\mathrm{\Pi}$, and it is the fact that $\Omega$ satisfies the $\varpi$-equation rather than $-\omega$ that distinguishes the two.

Within the framework of the $g_{ab}$ metric, we think of $\Omega$ as a kind of ``phantom field", whose role is simply to keep track of the \textit{physical} $\hat g_{ab}$ metric. It has energy tensor\footnote{See \cite{Newman}, footnote p.194 and \cite{Callan}. The latter call this quantity ``the new improved" energy tensor, for a conformally invariant massless scalar field.} 
$$
T_{ab}[\Omega] = \frac{1}{4\pi G} \; \Omega^3D_{ab}\Omega^{-1}
$$
the (appropriately conformally invariant) trace-free operator $D_{ab}$ being defined\footnote{See \cite{LeBrun,Eastwood}, and the footnote on p.124 of \cite{Penrose1984}.}  
$$
D_{ab} = \nabla_a\nabla_b - \frac{1}{4}\; g_{ab}\; \square - \frac{1}{2}\; R_{ab} + \frac{1}{8}\; Rg_{ab}.
$$

The Einstein equations (with $\Lambda$) for the $\hat g_{ab}$ metric, with massless sources having trace-free energy tensor 
$\hat T_{̂ab}$ can then be written in the remarkable form 
$$
T_{ab} = T_{ab}[\Omega]
$$
where $T_{ab}$ comes from the conformal scaling of the energy tensor for massless source fields:
$$
\hat T_{ab} = \Omega^{-2}\; T_{ab}.
$$

\textit{Every} solution of the $\varpi$-equation provides us with a conformal factor $\Omega(=\varpi)$ for which $R=12$, so to ensure uniqueness we need to impose something of the nature of two conditions per point on some spacelike initial 3-surface, here 
${\cal X}$, to ensure that we have the intended solution $\Omega$. A technical difficulty arises since $\Omega$ becomes infinite there, so it is easier to re-express the equation in terms of $\omega=-\Omega^{-1}$. Subtleties arise concerning the required two conditions per point, but it seems appropriate to adopt the suppressed rest-mass hypothesis, which asserts that in a power-series expansion
$$
\mathrm{\Pi}^a\mathrm{\Pi}_a = A + B\omega + C\omega^2 + D\omega^3 + \ldots
$$
we impose two conditions $B=0$, and $C=Q$, (in addition to the automatic $A=1$), throughout the 3-surface ${\cal X}$. Since the Yamabe ($\varpi$-)equation is second order of standard type, we might expect that this would be sufficient to ensure uniqueness. However, this is not quite the case, basically because of the following consideration. Let us contemplate the ordinary wave equation $\square\;\varphi=0$ in Minkowski space (with standard coordinates $t, x, y, z$). We get a unique solution for $\varphi$ if we specify both $\varphi$ and its $t$-derivative $\dot\varphi$ at all points of the ($t=0$) 3-plane ${\cal X}$. But if, instead, we specify the first $t$-derivative $\dot\varphi$ and \textit{second} $t$-derivative $\ddot\varphi$ on ${\cal X}$, this is clearly insufficient for uniqueness since solutions of the 3-dimensional Laplace equation $\nabla\psi=0$, in $x, y, z$ can be added to any solution $\varphi$ of $\square\;\varphi=0$ to get new solutions, without affecting either $\dot\varphi$ ̇or $\ddot\varphi$.

The situation with regard to the unique propagation from aeon to aeon in CCC appears to be similar, as can be seen from some general considerations: the specification of two local conditions of an invariant nature, such as $B=0$, and $C=Q$ on ${\cal X}$ will always fall somewhat short of providing complete uniqueness. To resolve this, we require an additional specification, like a boundary condition on a 2-surface at infinity within ${\cal X}$. This non-uniqueness is of a relatively mild character, of the nature of a symmetry breaking, and its resolution would appear to be related to issues referred to above concerning finding the appropriate anti-Higgs mechanism whereby rest-mass decays away in the very remote future of each aeon. This is a matter for future work on the mathematical underpinnings of CCC. For the moment, this relatively mild ambiguity of propagation from aeon to aeon will be ignored.

Because of the reciprocal hypothesis, the phantom field $\Omega$, when extended into $\check\Re$, has a completely different physical role to play from that in $\hat\Re$. It no longer appears as just a means of keeping track of the original physical 
$g_{ab}$ metric, but now it acts as a \textit{real} conformally invariant physical scalar field which satisfies the $\varpi$-equation in the $\check g_{ab}$ metric (by virtue of that equation's conformal invariance). This difference in interpretation comes about because $\Omega$'s conformal role is now the reverse of what it was before. It no longer simply keeps track of the physical metric ($\hat g_{ab}$) within the framework of the $g_{ab}$ metric, but it is now a physical field in its own right, taking up gravitational degrees of freedom from the previous aeon. Its satisfaction of the $\varpi$-equation (at least before rest-mass begins to appear) tells us that $\Omega$ initially is a self-coupled massless scalar. It is not one of the massless fields, such as electromagnetism, that would have been present in the late stages of the earlier aeon, contributing to its energy tensor $\hat T_{ab}$. These propagate directly through into $\check\Re$, giving an energy tensor $\check T_{ab}$conformally related to $\hat T_{ab}$ via 
$$
\hat T_{ab} = \Omega^{-2} T_{ab} \qquad \mathrm{and} \qquad \check T_{ab} = \omega^{-2} T_{ab}, \qquad \mathrm{so} \qquad 
\check T_{ab} = \Omega^4\hat T_{ab}.
$$
Instead, $\Omega$ provides a \textit{new} contribution to the total energy tensor that \textit{adds} to the contribution from $\check R_{ab}$. CCC interprets this new field as the initial form of newly created \textit{dark matter}\footnote{This material presumably has some relation to the ``Higgs field", since the appearance of mass in the dark matter would be associated with the appearance of mass generally, in the CCC scheme.} - so for the propagation of dark matter from aeon to aeon to maintain itself at the same level within each aeon, we require it to decay away completely in some way over the entire history of each aeon. In this initial form it is \textit{massless} but we find that as a result of the equations, the energy tensor $\check S_{ab}$ of the newly created material field\footnote{The quantity $\check S_{ab}$ referred to here is the \textit{sum} $\check V_{ab}+\check W_{ab}$ of \cite{Penrose2010}, subsection B11.} (in the $\check g_{ab}$ metric) must acquire a \textit{trace}, because we find \cite{Penrose2010}, subsection B10) that the equations tell us
$$
\frac{1}{12}\; R = 1 + (\Omega^2-1)^2\;(\mathrm{\Pi}^a\mathrm{\Pi}_a-1),
$$
rather than $R=12$, so the material involved in $\check S_{ab}$ has acquired a rest-mass, as indicated by 
$$
\check S_a^a = \frac{3}{2\pi G}\; (\Omega^2-1)^2\; (\mathrm{\Pi}^a\mathrm{\Pi}_a - 1) = \frac{3}{2\pi G}\; \omega^{-4}(\nabla^a\omega\nabla_a\omega - (1 - \omega^2)^2)
$$
and we note that the suppressed rest-mass hypothesis $\mathrm{\Pi}^a\mathrm{\Pi}_a - 1 = Q\omega^2+O(\omega^3)$ indeed suppresses the reappearance of rest-mass to a considerable degree - as philosophically desired by CCC. It should be pointed out that although $\check S_{ab}=O(\omega^{-2})$ and is still infinite at the Big Bang even with the suppressed rest-mass hypothesis, this hypothesis renders the rest-mass contribution to be very small in relation to the contributions from the other physical fields present at the Big Bang, since $\check T^b_a=O(\omega^{-4})$ (because $T_{ab}=O(1)$, so $\check T_{ab}=O(\omega^{-2})$ whence $\check T_{ac} \check g^{cb}=O(\omega^{-4})$).

Gravitational degrees of freedom in the remote future of the previous aeon, measured there by Weyl's conformal tensor $\hat C_{abcd}$, get converted, according to CCC, to derivatives of the initial dark matter field $\Omega$, just following our Big Bang. To see this, we need to understand the conformal behaviour of the gravitational field. Conformal invariance is most readily expressed in the 2-spinor notation \cite{Penrose1984}, and we use this notation (including \textit{abstract-indices}) here. The free-field wave equation for a massless particle of spin $\frac{1}{2}\; n(>0)$
$$
\nabla^{AA^\prime}\; \Phi_{ABC...E} = 0,
$$
$\Phi_{ABC...E}$ having $n$ totally symmetric 2-spinor indices 
$$
\Phi_{ABC...E} = \Phi_{(ABC...E)},
$$
is \textit{conformally invariant} under $\hat g_{ab}=\Omega^2\; g_{ab}$ when we rescale $\Phi_{ABC...E}$ by 
$$
\hat \Phi_{ABC...E} = \Omega^{-1}\; \Phi_{ABC...E}.
$$
When $n=2$ and $n=4$, respectively, we can relate 𝜙$\Phi_{ABC...E}$ to the Maxwell field tensor $F_{ab}$ and to the gravitational tensor $K_{abcd}$ referred to in Section 1 by
$$
F_{ab} = \Phi_{AB}\;\epsilon_{A^\prime B^\prime} + \bar \Phi_{A^\prime B^\prime}\; \epsilon_{AB} \qquad \mathrm{and} \qquad
K_{abcd} = \Phi_{ABCD}\; \epsilon_{A^\prime B^\prime}\; \epsilon_{C^\prime D^\prime} + \bar \Phi_{A^\prime B^\prime C^\prime D^\prime}\;\epsilon_{AB}\;\epsilon_{CD},
$$
where $\epsilon_{A^\prime B^\prime}$ and $\epsilon_{AB}$ are the skew-symmetrical ``epsilon spinors" used for lowering or raising spinor indices (and where $g_{ab}=\epsilon_{AB}\;\epsilon_{A^\prime B^\prime}$). The Weyl conformal tensor
$$
{C_{ab}}^{cd} = {R_{ab}}^{cd} - {2R_{[a}}^{[c}\; {g_{b]}}^{d]} + \frac{1}{3}\; R\; {g_{[a}}^{c}\; {g_{b]}}^{d},
$$
finds a similar expression
$$
C_{abcd} = \Psi_{ABCD}\epsilon_{A^\prime B^\prime}\epsilon_{C^\prime D^\prime} + \bar \Psi_{A^\prime B^\prime C^\prime D^\prime}\epsilon_{AB}\epsilon_{CD}
$$
($\Psi_{ABCD}$ being totally symmetric) and under conformal rescaling $\hat \Psi_{ABCD}=\Psi_{ABCD}$, which contrasts with the $\hat \Phi_{ABCD}=\Omega^{-1}\Phi_{ABCD}$ above, so that if we require $\hat \Phi_{ABCD}=\hat \Psi_{ABCD}$ then we find that 
$$
\Phi_{ABCD} = \Omega\; \Psi_{abcd}, \qquad \mathrm{i.e.}\; K_{abcd} = \Omega\; C_{abcd},
$$
as asserted in Section 1. The conformal invariance of $\Phi_{ABCD}$'s wave equation tells us that $K_{abcd}$ attains finite values at ${\cal X}$, whereas (since $\Omega\rightarrow\infty$ at ${\cal X}$) we find $C_{abcd}=0$ at ${\cal X}$. Since $C_{abcd}$ describes {\it R}'s conformal geometry at ${\cal X}$ it must be smooth across ${\cal X}$, vanishing at the Big Bang of
$\check\Re$, in accordance with the Weyl curvature hypothesis.

The \textit{information} in $K_{abcd}$ does, however, propagate into $\check\Re$ since the derivative in the normal direction to ${\cal X}$ (given by $\mathrm{\Pi}^a\nabla_a$) can be applied to $\omega K_{abcd}=-C_{abcd}$, giving 
$$
K_{abcd} = -\mathrm{\Pi}^e\; \nabla_e\;  C_{abcd} \qquad \mathrm{at} \; {\cal X}.
$$
Bianchi identities relate this normal derivative of $C_{abcd}$ to tangential derivatives of $R_{ab}$ at ${\cal X}$, so the gravitational degrees of freedom, given by $K_{abcd}$ at ${\cal X}$, are carried across ${\cal X}$, into the $g_{ab}$-Ricci tensor in $\check\Re$. Part of this information - in the ``magnetic part" of $K_{abcd}$ at ${\cal X}$ - is basically the Cotton(-York) tensor of the conformal 3-geometry of ${\cal X}$ itself (given by $\epsilon^{abcd}\; \mathrm{\Pi}_a\; \nabla_b\; R_{ce}$ at ${\cal X}$), while the ``electric part" of $K_{abcd}$ at ${\cal X}$ provides information propagating into $\check\Re^\prime$s interior.

The details of this are not essential for us. We see from general considerations that a gravitational wave within $\hat\Re$, propagating in some lightlike direction and impinging upon ${\cal X}$ from its past will create a disturbance to the future of ${\cal X}$ continuing within $\check\Re$ in that same direction, as is clear when we refer things to the $g_{ab}$ metric, since the wave equation for $K_{abcd}$ (i.e. for $\Phi_{ABCD}$) simply carries $K_{abcd}$ across ${\cal X}$ without obstruction. Converting to the $\check g_{ab}$ metric, we do not see this information initially in $\check K_{abcd}$ owing to its suppression by the factor $\omega^2$ in $\check\Re$ and the gravitational degrees of freedom are transferred to higher derivatives of $\Omega$. The key point is that no information can be lost because the passing from $g_{ab}$ to $\check g_{ab}$is reversible. An impulsive $K_{abcd}$ disturbance in the $g_{ab}$ description continues as a disturbance in the $\Omega$-field when viewed from the $g_{ab}$ perspective, so it must do so also when viewed from the $\check g_{ab}$ perspective.

\section*{Appendix B: CCC explanation of low temperature-variance rings}

According to CCC, a major contribution to the CMB temperature variation comes from black-hole encounters in the aeon prior to ours, each producing a huge, effectively impulsive, basically isotropic, burst of gravitational radiation in the previous aeon. Upon reaching the last-scattering 3-surface ${\cal L}$ of our aeon, it would be seen as a circular ring of either slightly raised or slightly lowered overall temperature, largely uniform over the ring. Such a burst, if early enough in that aeon, would be likely to intersect many other such bursts from other sources in the previous aeon. The picture presented in our CMB sky is of numerous intersecting circles (like ripples on a pond during a burst of rain), each circular disturbance carrying a slightly warmer or cooler temperature than average (depending on geometry described in Section 1), this temperature being basically uniform around the circle, but altered at each intersection point with another such circle (Fig. 5). In the CCC picture, these circular disturbances should be the major effect of relevance to the small scale under consideration.

\begin{figure}\sidecaption
\resizebox{0.35\hsize}{!}{\includegraphics*{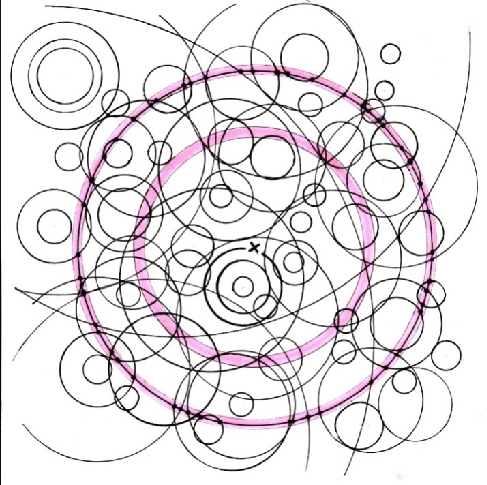}}
\caption{Schematic picture of CCC's proposed small-scale structure of temperature variations in the CMB, where numerous circles of either higher or lower temperature, each distributed fairly uniformly about its circle, all contribute to the total effect. In the procedure for locating them, a central point (here marked ``x") is selected and the temperature variance is determined over successive rings with that centre. When a ring does not contain such a circle the main variance is obtained through its intersections with other such circles, but when it does, there is a dominant contribution from that circle, and the variance is reduced through non-linear (averaging) effects at its intersections with other circles.}
\end{figure}
 
In our procedure for detecting such circular disturbances, we choose some point $\times$ in the CMB sky as a prospective centre, and then determine the temperature variance over circular rings centred about $\times$ (of annular width $0.5^\circ$ and angular radius increasing in steps of $0.5^\circ$) as a function of angular radius $r$. For a randomly chosen such ring, the major contribution to its variance, according to the above, would come from its intersection points with the circles of disturbance centred at other places, and the temperature might be raised or lowered at these points (Fig.5). When the value of $r$, for given $\times$, happens to produce a ring that \textit{contains} one of the circles of disturbance, then we find that this disturbance contributes a comparatively large additional effect very uniformly all around the ring, this dominating the temperature as we move around the ring.

Had it been the case that the observed temperature at any particular point on the ring came from simply \textit{adding} together, at that point, the effects due to all the circular disturbances that pass through that point, then the result of the \textit{variance} calculation would not be sensitive to whether or not the chosen ring might happen to contain one of the sought-for circular disturbances. The temperature along the ring would simply have that ring's constant value added to what it would have been otherwise. The \textit{mean} value would be raised or lowered, but the variance unaffected. However, it turns out that CCC provides an important \textit{non}-additive effect, and this is what we need to examine.

Consider the history of the disturbance from a black-hole event under consideration. In the previous aeon, these are gravitational-wave disturbances, and many of these will cross many other such waves before ${\cal X}$ is reached. However, non-linear gravitational effects, despite the huge energies involved, would provide only a tiny contribution to the essentially linear propagation of the waves, linearized gravitational theory being adequate, the resulting effects being indeed additive. Moreover, even as the waves cross ${\cal X}$ over into the present aeon, linear propagation is \textit{still} expected, initially, according to the equations of CCC. The reason, as explained in Appendix A, is that the deliberate choice of propagation equations in $\check\Re$, for the \textit{initial} behaviour just following the Big Bang, is that they simply continue the propagation in $\hat\Re$ across ${\cal X}$, without changing the equations of motion, but merely interchanging the conformal factors $\Omega$ and $-\omega$, thereby ensuring consistency of the equations.

However, when mass begins to enter the picture (presumably around the moment where the universe cools to around the ``Higgs temperature") we must expect that the initial dark matter begins to acquire \textit{viscosity}, owing to its newly acquired rest-mass. We now regard the disturbances as leading to fluid motions in the early dark matter, and when two of our rings intersect, their motions would tend to \textit{average out} rather than their temperatures simply adding up. Such non-linearities lead to a lowering of variance, rather than simply changing the mean and leaving the variance alone, as the following simple consideration shows.

Suppose we have random variables $x_1, x_2,..., x_n$; then adding a constant a to each will clearly not affect the variance, as measured by the standard deviation $\sigma$ 
$$
\sigma^2 = \frac{1}{n}\; \sum^n_{i=1}(x_i - \bar x)^2 = \frac{1}{n}\;\sum^n_{i=1}((x_i+a)-(\bar x+a))^2
$$
whereas \textit{averaging} each with the constant $a$ will reduce $\sigma$ by a factor of 2 
$$
\frac{1}{n}\sum^n_{i=1}\;(\frac{1}{2}(x_i+a)-\frac{1}{2}(\bar x+a))^2 = (\frac{1}{2}\;\sigma)^2.
$$
It should be borne in mind, however, that this averaging, as opposed to additivity, occurs only for the ring intersections lying within the region between ${\cal X}$ and ${\cal L}$, so the variance reduction would be expected to be much smaller than this factor of 2 Moreover complicating matters, such as energy generated in the frictional aspects of this viscosity, might need to be taken into consideration. Nevertheless, it is a reasonable expectation that a sizable variance reduction will indeed occur just at the point when the radius $r$ reaches a value where the ring contains the circle of disturbance, and this is what we appear to be seeing.

\section*{Acknowledgements}

We thank A.Ashtekar, and the Institute for Gravitation and the Cosmos, for financial support, E.T.Newman, K.P.Tod, P.J.E.Peebles, P.Ferreira, M.Wisse, B.Schutz, K.Meissner and L.Smolin for valuable discussions, A.L.Kashin and H.Khachatryan for help with data, and the referee for useful comments. We gratefully acknowledge the use of data of WMAP, \textit{lambda.gsfc.nasa.gov}.

\end{document}